\documentclass[preprint]{revtex4-1}
 
\usepackage{graphics}
\usepackage{graphicx}
\usepackage{pstricks}
\usepackage{amsbsy}

\renewcommand{\vec}{\mathbf}

\newcommand{\vecs}{\boldsymbol}

\renewcommand{\d}{\mathrm{d}}
\newcommand{\w}{\widetilde}
\newcommand{\ext}{\mathrm{ext}}

\begin{document}

\title{Multiscale response of ionic systems to a spatially varying
  electric field}

\author{J. S. Hansen} \email{jschmidt@ruc.dk}
\affiliation{
  ``Glass and Time'', IMFUFA, Department of Science and Environment,
  Roskilde University, Postbox 260, DK-4000 Roskilde, Denmark 
}

\begin{abstract}
  In this paper the response of ionic systems subjected to a spatially
  varying electric field is studied. Following the Nernst-Planck
  equation, two forces driving the mass flux are present, namely, the
  concentration gradient and the electric potential gradient. The mass
  flux due to the concentration gradient is modelled through Fick's
  law, and a new constitutive relation for the mass flux due to the
  potential gradient is proposed. In the regime of low screening the
  response function due to the potential gradient is closely related
  to the ionic conductivity. In the large screening regime, on the
  other hand, the response function is governed by the charge-charge
  structure. Molecular dynamics simulations are conducted and the two
  wavevector dependent response functions are evaluated for models of
  a molten salt and an ionic liquid. In the low screening regime the
  response functions show same wavevector dependency, indicating that
  it is the same underlying physical processes that govern the
  response. In the screening regime the wavevector dependency is very
  different and, thus, the overall response is determined by different
  processes. This is in agreement with the observed failure of the
  Nernst-Einstein relation.
\end{abstract}

\maketitle

\section{Introduction}
Ionic liquids, molten salts and ionic solutions show a response to the
application of an external electric field. In the case of small field
amplitudes the response is typically modelled by linear constitutive
relations, for example, the response manifested by a charge current is
related to the local field by Ohm's law and the charge density to the
external field through the charge-charge correlation function
\cite{hansen:pra:1975,hansen:book:2006}. The response is characterized
by different response functions (or transport coefficients) like the
electric conductivity, electric permittivity and the charge-charge
correlation. One can often find relations between the different
response functions \cite{hansen:book:2006}; at least in some limits. A
more famous one is the Nernst-Einstein equation that relates the self
diffusion coefficient to the electric conductivity
\cite{smedley:book:1984}, i.e., single particle flux to the charge
current. This is a quite surprising relation as the particle flux is a
single particle phenomenon whereas the charge current is a collective
phenomenon. The Nernst-Einstein equation is then only valid when the
ion cross-correlations can be neglected
\cite{hansen:pra:1975,harris:jpcb:2010}. One example where this
assumption is not valid is where the flux of ion-pairs contributes to
the mass flux, but not to the charge current as the charges cancels
\cite{hansen:book:2006}. The deviation can be determined from
simulations or experiments and is often quantified by a deviation
parameter \cite{hansen:book:2006}, which, interestingly, Harris et
al. \cite{kanakubo:jcpb:2007,harris:jpcb:2008} have expressed in terms
of the velocity cross-correlation functions.  Importantly, the failure
of the Nernst-Einstein equation means that the particle flux due to
the electric field cannot be modelled through Ohm's law
directly. Rather than approaching this problem through the deviation
parameter it is appealing to take one step back and propose a linear
constitutive relation that involves a new response function relating
the mass flux to the external field directly. This is done in this
paper.
 
The system's response is dependent on the wavelength of the external
field, and this can be modelled through wavevector dependent response
functions \cite{evans:book:2008,hansen:book:2006}. Investigating the
wavevector dependence is relevant as the response can vary as function
of length scale \cite{hansen:langmuir:2015}. Also, this provide a way
to probe a characteristic correlation lengths for a given system
\cite{hansen:langmuir:2015, furukawa:prl:2009}; if the characteristic
length scales are different for the different response functions this
indicates that different physical underlying mechanisms are
responsible for the system response. This is also addressed here.

The paper is organized as follows: In the next section the theory for
the response of an ionic system subjected to a static sinusoidal
external field is presented. In Sect. \ref{sec:III} molecular dynamics
simulation results are presented and discussed, and, finally, in the
last section conclusions from the work are drawn.

\section{Theory}
We consider an ionic system composed of one cation and one anion
specie. The ions are rigid meaning that any higher order induced effects
and electron transfer mechanisms are ignored. The charges are $\pm
q$, respectively.  Let $i$ indicate either a cation or an anion, i.e.,
$i=+$ or $-$, then the number density $n_i$ follows the balance
equation \cite{degroot:book:1984}
\begin{equation}
\label{eq:balance}
\frac{\partial n_i}{\partial t} = \sigma_i - \vecs{\nabla}\cdot 
n_i\vec{c}_i - \vecs{\nabla}\cdot n_i\vec{u} \, ,
\end{equation}
where $n_i\vec{c}_i$ is the diffusive flux and $n_i\vec{u}$ the
advective flux. The production term $\sigma_i$ accounts for additional
forces that generate a local change in $n_i$; this includes
application of an external electric field. The terms on the right-hand
side of Eq. (\ref{eq:balance}) can be expressed as the divergence of
fluxes such that if one writes the production term as $\sigma_i =
-\vecs{\nabla} \cdot \vec{j}_i^e$ and $n_i\vec{c}_i = \vec{j}_i^d$ we
have for zero advection
\begin{equation}
\frac{\partial n_i}{\partial t} = - \vecs{\nabla}\cdot \vec{j}_i = -
\vecs{\nabla}\cdot(\vec{j}_i^{e} + \vec{j}_i^{d}) \, . 
\label{eq:massbalanceflux}
\end{equation}
The system is kept away from equilibrium by application of a static
spatially varying external electric field. The field points and varies
in the direction parallel to the system $z$-direction, i.e, the
non-zero $z$-component of the external field reads
\begin{equation}
\label{eq:extfield}
E_z^\ext(z) = E_0 k_n^m\cos(k_n z) \, ,
\end{equation}
where $k_n = 2\pi n/L_z$ is the wavevector, $n=1, 2, \ldots$, and
$L_z$ is the system length in the $z$-direction. $m$ is either 0 or
1. The experimental realization of this field is not
straightforward. Here it is considered as we are interested in the
wavevector dependent response and as such this resembles the
sinusoidal transverse and longitudinal force field methods (STF and
SLF), see for example
Refs. \onlinecite{baranyai:pra:1992,hoang:jcp:2012,dalton:pre:2013}.
The corresponding electric potential is
\begin{equation}
\phi^\ext(z) = - \int_0^z E_z^\ext(z') \, \d z' + \phi^\ext(0) = - E_0
k_n^{m-1}\sin(k_n z) \, ,
\end{equation} 
using $\phi^\ext(0)=0$.  Note, $m=0$ corresponds to a wavevector
independent field amplitude and $m=1$ to wavevector independent
potential amplitude.

It is in place to discuss the Maxwell equations. First, the
induced/screening field is $E=E(z)$ and according to Gauss' law $\d
E/\d z = \rho_q/\epsilon_0$, where $\rho_q$ is the charge density
given by the induced ionic density, $\epsilon_0$ is the electric
permittivity of free space.  From the Maxwell-Faraday equation
$\vecs{\nabla}\times \vec{E} = -\dot{\vec{B}} = \vec{0}$, that is, the
field due to the screeing does not result in any change in the magnetic
field $\vec{B}$. Then Gauss' law for the magnetic field is fulfilled,
$\vecs{\nabla} \cdot \vec{B} = 0$. Furthermore, since $\vecs{\nabla}
\times \vec{B}= \vec{0}$ and $\dot{\vec{E}}=\vec{0}$ there are no net
charge current (Ampere's circuital law). The system is therefore in a
steady state.

To proceed one needs to relate the fluxes with the corresponding
forces \cite{degroot:book:1984}. For sufficiently small force
amplitude this is done through the generalized linear response
theory. Consider the mass flux in the $z$-direction $j_i$ to depend on $N$
forces $X_n$, $n=1,2,\ldots N$ then we have in the homogeneous
situation
\begin{equation}
j_i = -\sum_{n} \int_0^\infty \int_{-\infty}^\infty
\chi'_{n}(\vec{r}-\vec{r}',t-t') X_n(\vec{r}',t') \, \d \vec{r}' \d t'  \, ,
\end{equation}
where $\chi'_n$ is the response function relating the flux $j_i$ to
the force $X_n$. \color{red}Since the system is in a steady state we
can safely ignore time memory effects and, furthermore, assuming
isotropy the response functions \color{black} can then be written as
$\chi'_n(\vec{r}-\vec{r}',t-t')= \chi_n(z-z')\delta(t-t')$. The flux
is
\begin{eqnarray}
  j_i &=& -\sum_n \int_0^\infty \delta(t-t') \int_{-\infty}^\infty
\chi_{n}(z-z') X_n(z',t') \, \d z' \d t'  
\nonumber \\
&=& - \sum_n \int_{-\infty}^\infty \chi_{n}(z-z') X_n(z',t) \, \d z'
= - \sum_n \int_{-\infty}^\infty \chi_n(z-z')X_n(z') \, \d z'
\end{eqnarray}
The final expression is due to \color{red} the steady state conditon \color{black}.
This generalized response formalism can be applied to the present
situation. The flux is proposed to be given by the two terms
$(N=2)$ 
\begin{equation}
j_{i} = j_i^d + j_i^e = -\int_{-\infty}^\infty D_i(z-z') \frac{\d n_i}{\d z'}
\, \d z' - \frac{1}{q_i}\int_{-\infty}^\infty \chi_i(z-z') \frac{\d
  \phi^\ext}{\d z'} \, \d z' \, ,
\label{eq:massflux}
\end{equation}
where $D_i$ is the diffusion response function (or diffusion
coefficient) and $\chi_i$ is the response function that relates
the \color{red}\emph{mass} flux to the external
field. \color{black} The first relation is simply a generalized
version of Fick's law, but the second relation is not a generalization
of Ohm's law as $\chi_i$ relates the mass flux directly to external
electric potential. Also, the force in Ohm's law is given by the
\color{red}\emph{local} \color{black} electric potential, i.e., the
sum of the screening potential and the external potential. The
$\chi_i$-response function can be interpreted as the system response
to an external field excluding the effects from diffusion. Note that
Eq. (\ref{eq:massflux}) is a generalized form of the Nernst-Planck
equation \cite{bruus:book:2008}.

In the steady state $j_i^d +j_i^e = 0$, and one has
\begin{equation}
  \int_{-\infty}^\infty D_i(z-z')\frac{\d n_i}{\d z'} \, \d z' =
  -\frac{1}{q_i} \int_{-\infty}^\infty \chi_i(z-z') \frac{\d
    \phi^\ext} {\d z'} \d z' \, .
\end{equation}
In Fourier space by the convolution theorem for wavevector
$\vec{k}=(0,0,k_n)$ this reads 
\begin{equation}
  i k_n \w{D}_i(k_n) \w{n}_i(k_n) = - \frac{i k_n
  }{q_i}\w{\chi}_i(k_n)\w{\phi}^\ext(k_n)  
\end{equation}
or
\begin{equation}
  \w{n}_i(k_n)=-\frac{\w{\chi}_i(k_n)}{q_i\w{D}_i(k_n)}
  \w{\phi}^\ext(k_n) \, . \label{eq:FourCompRho}
\label{eq:fouriercomponets}
\end{equation}
Equation (\ref{eq:fouriercomponets}) is the expression for the Fourier
coefficients for the number density. \color{red}{For $\w{\phi}^\ext(k)
  >0$ the Fourier component for the cation and anion are negative and
  positive, respectively.} \color{black} From this result one can
also find the Fourier coefficients for the charge density,
$\w{\rho}_q$. First, it is observed that due to symmetry the number
density follows a sine series, i.e.,
\begin{equation}
n_i(z) = n_{0} + \sum_{j=n}^\infty \w{n}_{i,j}(k_j)\sin(k_j z)
\, . \label{eq:multimode}
\end{equation}  
The Fourier components of the charge density is then
\begin{equation}
\label{eq:FourierCharge}
\w{\rho}_q(k_n) = q_+ \w{n}_+ + q_- \w{n}_- =
-\left(\frac{\w{\chi}_+(k_n)}{\w{D}_+(k_n)} +
  \frac{\w{\chi}_-(k_n)}{\w{D}_-(k_n)}\right) \w{\phi}^\ext(k_n) \, .
\end{equation}
For small field strengths and negligible screening only the
fundamental mode $k_n=2\pi n/L$ is excited and we have that
\begin{equation}
  n_i(z) \approx n_{0} +
  \w{n}_{i}(k_n) \sin(k_n z) \, .
  \label{eq:FourierChargeSmall}
\end{equation} 

\color{red} In the following, focus is on the case where
Eq. (\ref{eq:FourierChargeSmall}) is true \color{black} and where the
two ionic species, $+$ and $-$, have same transport properties $\chi_i
= \chi$, and $D_i=D$. Then Eq. (\ref{eq:FourierCharge}) reduces to
\begin{equation}
\label{eq:FourierCompCharge}
\w{\rho}_q(k_n) = -\frac{2\w{\chi}(k_n)}{\w{D}(k_n)}\w{\phi}^\ext(k_n) \, .
\end{equation}
From linear response theory \cite{hansen:book:2006} the Fourier
components for the charge density is related to the charge-charge
correlation function (or charge-charge structure) $S_{ZZ}$ by
\begin{equation}
\label{eq:linresponse}
\w{\rho}_q(k) = -\frac{nS_{ZZ}(k)}{k_B T} \w{\phi}^\ext(k) \, ,
\end{equation}
where $n=n_+ + n_-$.  We then have an expression for $\w{\chi}$ in
terms of the diffusion coefficient and the charge-charge structure
\begin{equation}
  \w{\chi}(k_n) = \frac{n \w{D}(k_n)}{2k_BT}S_{ZZ}(k_n) \, .
\label{eq:endeq}
\end{equation}
The charge-charge structure is a collective property, and from
Eq. (\ref{eq:endeq}) one can see that $\chi$ relates this collective
property to the single particle property governed by the diffusion
coefficient. 

It is worth noting that in the Debye-H\"{u}ckel regime, $k_BT \gg
q\phi$, the charge-charge correlations are negligible, i.e., $S_{ZZ}(k) =
1$. This corresponds to the limit of zero screening \color{red}{and a
  relative permittivity of unity}. \color{black} Equation (\ref{eq:endeq}) then
reads
\begin{equation}
\w{\chi}(k) = \frac{n \w{D}(k)}{2k_BT} \, , \ \ \ \ \ 
\text{(Debye-H\"{u}ckel regime)}
\end{equation}
which is equivalent to the Nernst-Einstein equation
\cite{smedley:book:1984} and $\chi$ can in this limit be interpreted
as the ionic electric conductivity. The charge density Fourier
components are in this limit $\w{\rho}_q = -n\w{\phi}^\ext/k_BT $,
i.e., they only dependent on amplitude of the external field. For
systems where the diffusion coefficient is wavevector independent,
$\w{D}(k)\approx D_0$, the response function is
\begin{equation}
  \label{eq:screening}
\w{\chi}(k) = \frac{nD_0}{2k_BT}S_{zz}(k) \,
. \ \ \ \ \ (\text{Screening regime})
\end{equation}
This means that the wavevector dependent response in the presence of
an external electric field is dominated by the screening effects.

\section{Simulations and Results \label{sec:III}}

\subsection{Simulation details}
The response is evaluated for two simple models: (i) one model for
molten salt proposed by Hansen and McDonald \cite{hansen:pra:1975} and
(ii) one \color{red}{modified }\color{black} model for ionic liquids
used by Chapela et al. \cite{chapela:jcp:2015}. For the molten salt
the ions are simple spherical particles with same mass and point
charges $\pm q$. The van der Waals interaction is the inverse power
law function $V(r) = \epsilon (\sigma/r)^9$, where $r$ is the distance
between two ions, $\epsilon$ and $\sigma$ define the energy and length
scale, respectively. The Coulomb interactions are calculated through
the shifted force method \cite{fennell:jcp:2006,hansen:jpcb:2012},
$\vec{F}(r) = q_iq_j(1/r^2 - 1/r_c^2)\vec{r}/r$, for $r\leq r_c$. Here
$\vec{r}$ is the vector of separation with magnitude $r$, and $r_c$ is
the cut-off radius set to $r_c = 3 \sigma$; this cut-off distance is
also used for the van der Waals interactions. The positions of the
particles are integrated forward in time with the leap-frog algorithm
\cite{frenkel:book:1996} and the temperature is controlled using a
Nos\'{e}-Hoover thermostat
\cite{nose:molphys:1984,hoover:pra:1985}. In all simulations the total
ion number density is $n = 0.368 \sigma^{-3}$; the number of ions are
1000, giving 500 ion-pairs. Two different temperatures are simulated,
$T=0.0177 \epsilon/k_B$ and $1.0177 \epsilon/k_B$, the former being a
realistic temperature for the model. To simulate the Debye-H\"{u}ckel
regime $k_BT \gg q\phi$ the ion-ion Coulomb interactions are removed
whilst keeping the temperature fixed at $T=1.0177 \epsilon/k_B$; this
system is symbolized using $T_\infty$. \color{red}{Alternatively, one
  can perform simulations at very high temperatures, but this will
  result in numerical instabilities.} \color{black} In the following
all quantities are given in units of $\sigma$, $\epsilon$, $q$, and
mass $m$, and as it is common practise these the units are not written
explicitly.

\color{red}{For the simple molten salt system the shifted force method
  can be tested against the direct Ewald summation method
  \cite{toukmaji:cpc:1996}. From equilibrium simulations it was found
  from the structure that the Ewald method converges satisfactory
  using 124 replica systems and that it agrees with the data from the
  shifted force method, see also
  Ref. \onlinecite{hansen:jpcb:2012}. For the non-equilibrium
  situation at $T=1.0177$ the Ewald and shifted force methods yield
  same results for all wavevectors tested $0<k<2.2$.}\color{black}

The modified ionic liquid model is composed of cations with a
spherical point charge particle (head group) and two spherical
non-charged tail particles. The particles in the cation are
linearly connected using a simple spring force
$\vec{F}=-k(r-1)\vec{r}/r$, where $k=100$ is the spring constant.  
Anions are simple spherical point charge particles \cite{chapela:jcp:2015}. 
\color{red}{ Rather than a hard-sphere type potential in the original
  model, the van der Waals interactions are here given through the
  Weeks-Chandler-Andersen potential \cite{weeks:jcp:1971} $V(r) =
  4((1/r)^{12} - (1/r)^6)$, where the cut-off is set
  at $r_c=2^{1/6}$.The Coulomb interaction is given by the
  Yukawa potential $V(r)=q^2 e^{-r/\lambda_D}/r$, with $\lambda_D=1/2$
  corresponding to a relative small Debye screening length and
  the reduced charge is $q=4$. The cut-off distance for the Yukawa
  potential is set to $r_c=2.5$. The state point is $(n,T)=(1,1)$ and
  the simulation method is the same as for the molten salt
  simulations. This choice of parameters gives, qualitatively, the
  fluid structure observed in different ionic liquid
  \cite{hayes:cr:2015,sanchez-badillo:jpcb:2015}.  }\color{black}The
number of particles are 864, that is, 216 ion pairs.

Simulations of the non-equilibrium system is also performed. Here an
additional force from the external field, Eq. (\ref{eq:extfield}), is
added to the total force experienced by the ions $\vec{F}_i^\ext =
q_iE^\ext \vec{k}$, where $\vec{k}$ is the unit vector parallel to the
$z$-axis.

\subsection{Results: Molten salt \label{sec:moltensalt}} 
\begin{center}
  \begin{figure}
    \includegraphics[scale=0.3]{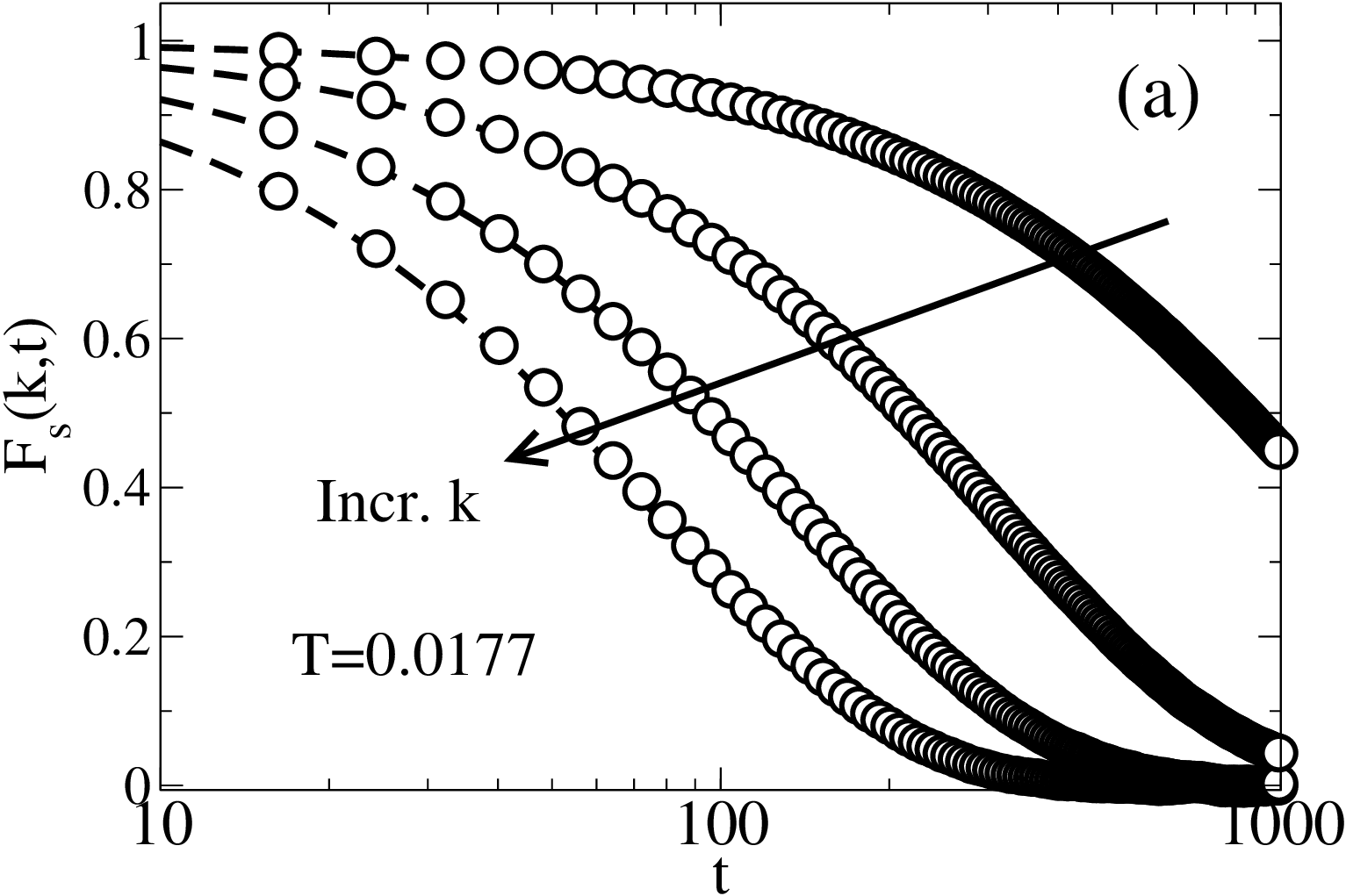}
    \includegraphics[scale=0.3]{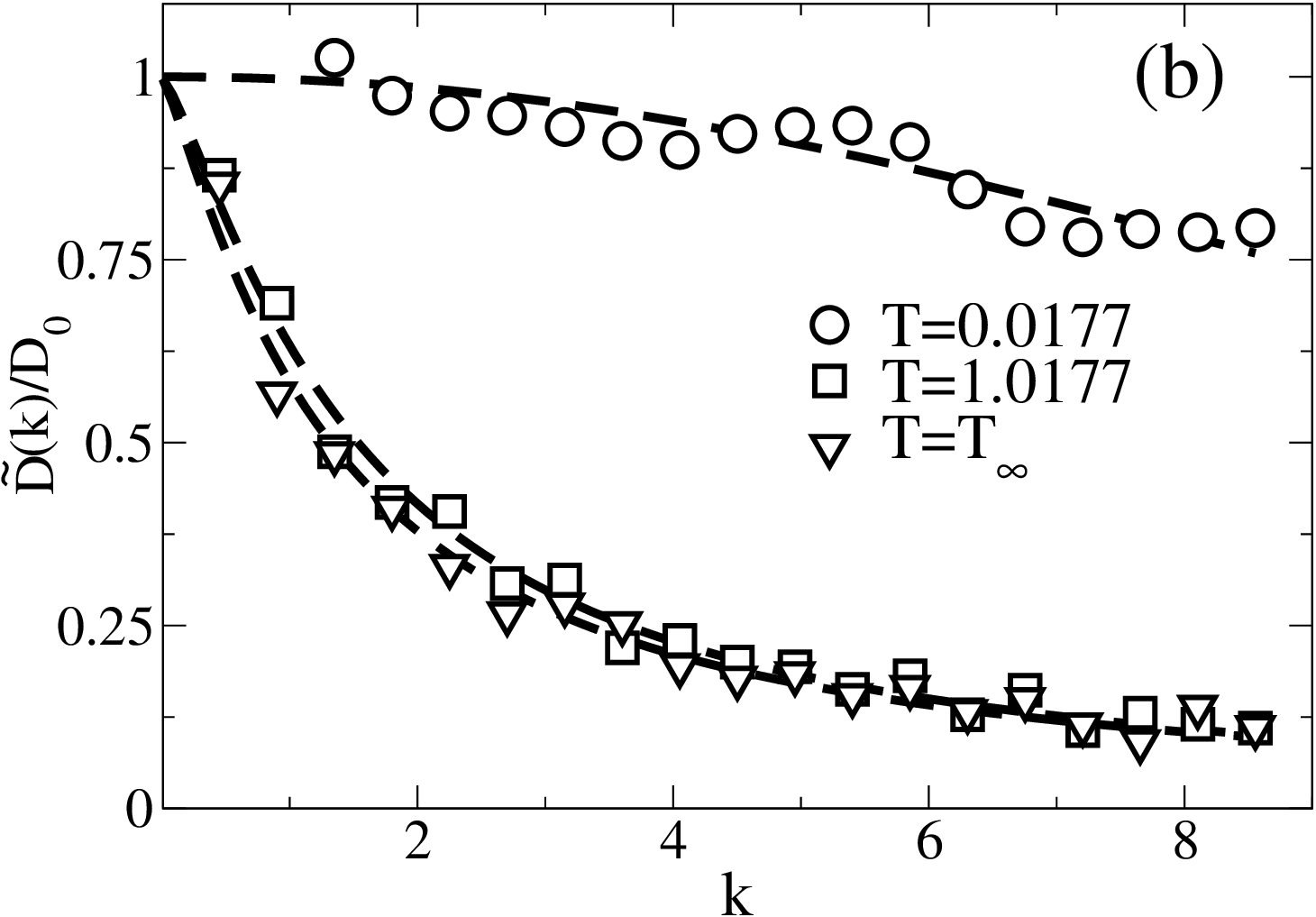}
    \caption{ \label{fig:scatteringDkernel} Molten salt (a):
      Incoherent intermediate scattering function for different 
      wavevectors (circles). Punctured line is $f(k,t) =
      e^{-\frac{1}{6}\langle \Delta r^2 \rangle k^2 t}$, where
      $\langle \Delta r^2\rangle$ is the mean square displacement.
      (b) The diffusion kernel at different temperatures. Punctured
      lines are best fit to Eq.(\ref{eq:lorentz}). 
      \color{red}Parameter values are for $T=0.0177, 1.0177$ and
      $T_\infty$, respectively: $D_0=0.011, 0.84, 0.92$,
      $\alpha=0.0073, 0.69, 0.78$, and $\beta=1.72, 1.14,
      1.12$.\color{black}} 
  \end{figure}
\end{center}

The wavevector dependent diffusivity can be obtained as follows. The
Gaussian approximation \cite{hansen:book:2006, boon:book:1991} relates the
diffusion coefficient to the incoherent intermediate scattering
function (or the self-part of the density-density correlations), so in
the diffusive regime, i.e., for large $t$, this is here generalized to
\begin{equation}
  F_s(k,t)=e^{-\w{D}(k) k^2 \, t}  \, .
  \label{eq:gauss}
\end{equation}
The Fourier-Laplace transformation is 
\begin{equation}
  S_s(k, \omega) = \int_0^\infty e^{-i\omega t} e^{-\w{D}(k)k^2
    t} \, \d t = \frac{1}{i\omega + \w{D}(k) k^2}  \, ,
\end{equation}
which gives an expression for the wavevector dependent diffusivity in
the limit of zero frequency
\begin{equation}
  \w{D}(k) = \frac{1}{k^2 S_s(k,0)}  \, .
  \label{eq:GaussKernel}
\end{equation}
Microscopically the intermediate scattering function is defined from
the ensemble average
\cite{boon:book:1991}
\begin{equation}
  F_s(k,t)=\frac{1}{N}\left \langle \sum_i e^{-ik(z_i(t)-z_i(0))}
  \right \rangle \, 
\end{equation}
where $N$ is the number of ions and is thus a single particle
property.  In Fig. \ref{fig:scatteringDkernel} (a) the intermediate
scattering function is plotted for different wavevectors in the case
of $T=0.0177$. Also, shown as punctured lines $f(k,t) =
e^{-\frac{1}{6}\langle \Delta r^2 \rangle k^2 t}$, where $\langle
\Delta r^2\rangle$ is the particle mean square displacement. It is
seen that the Gaussian approximation holds surprisingly well for this
model validating Eq.(\ref{eq:gauss}). The data are Fourier-Laplace
transformed and the Gaussian diffusion kernel is found from
Eq. (\ref{eq:GaussKernel}); the results are plotted in
Fig. \ref{fig:scatteringDkernel} (b). The function
\begin{equation}
\label{eq:lorentz}
\w{D}(k)=D_0/(1 + \alpha k^\beta) 
\end{equation}
is fitted to data where the zero wavevector diffusion coefficient,
$D_0$, is found from the mean square displacement $\langle \Delta
r^2\rangle = 2 D_0 t$ for $t\rightarrow \infty$. It is observed that
the normalized kernel is identical for the two cases $T=1.0177$ and
$T=T_\infty$. For $T=0.0177$ the diffusivity features a relative low
wavevector dependency in the range studied here and we have
$\w{\chi}(k) \propto S_{ZZ}(k)$ according to Eq. (\ref{eq:screening}).

\begin{center}
  \begin{figure}
    \includegraphics[scale=0.3]{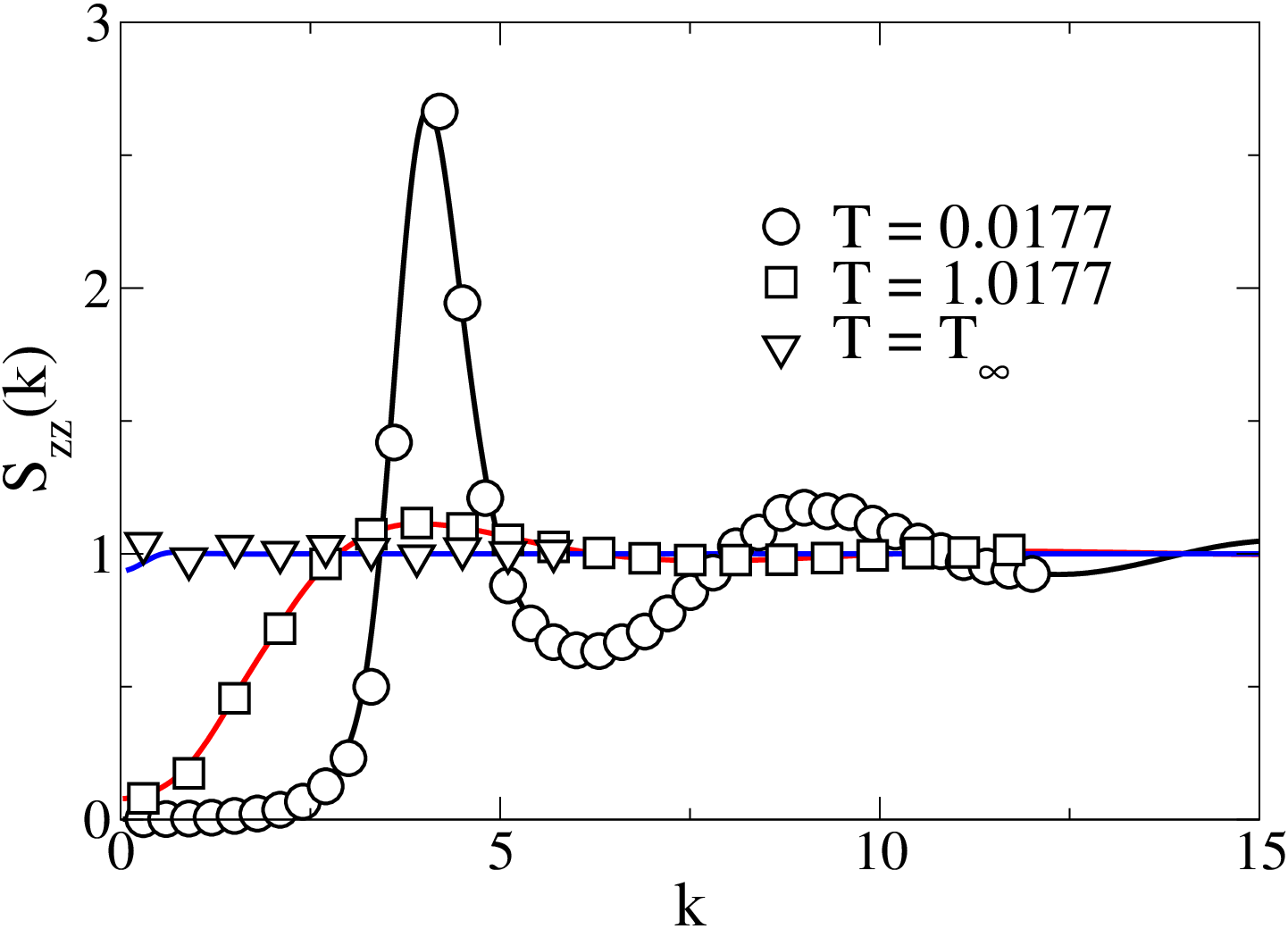}
    \caption{ \label{fig:Szz}
      Charge-Charge structure for the molten salt model. Symbols are
      data obtained from Eq. (\ref{eq:SzzDirect}) and full lines are from
      Eq. (\ref{eq:SzzRadial}) 
    }
  \end{figure}
\end{center}

Next the charge-charge structure is evaluated. This is defined as
\cite{hansen:book:2006}  
\begin{equation}
S_{ZZ}(\vec{k}) = \frac{1}{N}\left \langle
\rho_q(\vec{k},0) \rho_q(-\vec{k},0) 
\right\rangle \, ,
\label{eq:SzzDirect}
\end{equation}
where $\rho_q(\vec{k},0)=\sum_i q_i
e^{-i\vec{k}\cdot\vec{r}_i}$. Note, this is a collective property. For
non-zero wavevectors $S_{zz}(k)$ can also be calculated from the
radial distribution functions, see
e.g. Ref. \onlinecite{hansen:pra:1975},
\begin{equation}
  S_{zz}(k) = 1 + \frac{2\pi n}{k}\int_0^\infty \Delta g(r) r\sin(k r)
  \, \d r \, ,
\label{eq:SzzRadial}
\end{equation}
where $\Delta g(r)$ is the difference between the cation-cation and
cation-anion radial distribution functions, $\Delta g(r) = g_{++}(r) -
g_{+-}(r)$. The charge-charge structure is plotted in
Fig. \ref{fig:Szz} for the three different systems; symbols are data
from Eq. (\ref{eq:SzzDirect}) and lines are $S_{zz}$ calculated from
Eq. (\ref{eq:SzzRadial}). As expected we observe
a zero screening, $S_{zz}=1$, for $T=T_\infty$, but
non-negligible screening for $T=1.0177$ and $T=0.0177$.

\begin{figure}
  \includegraphics[scale=0.3]{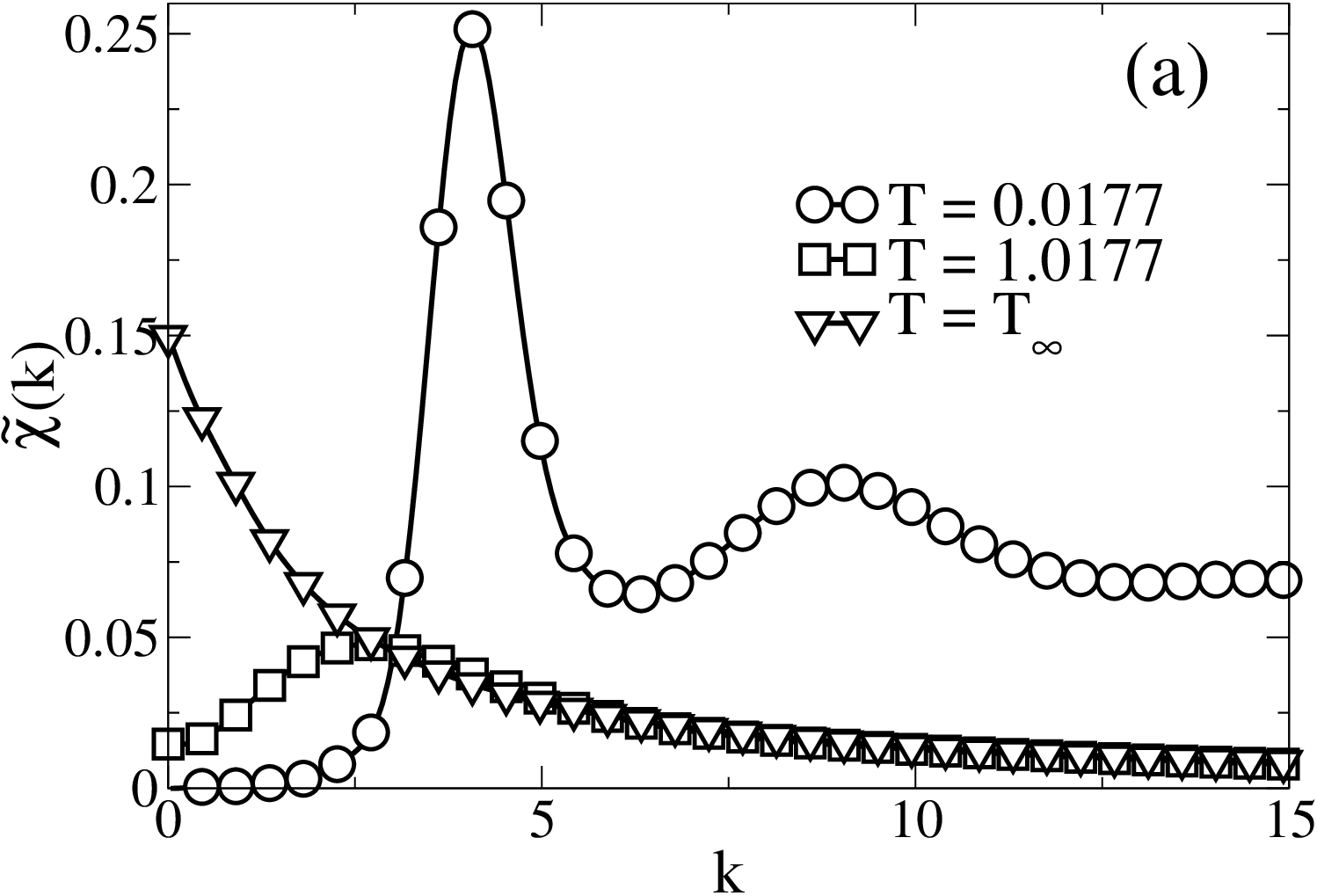}
  \includegraphics[scale=0.3]{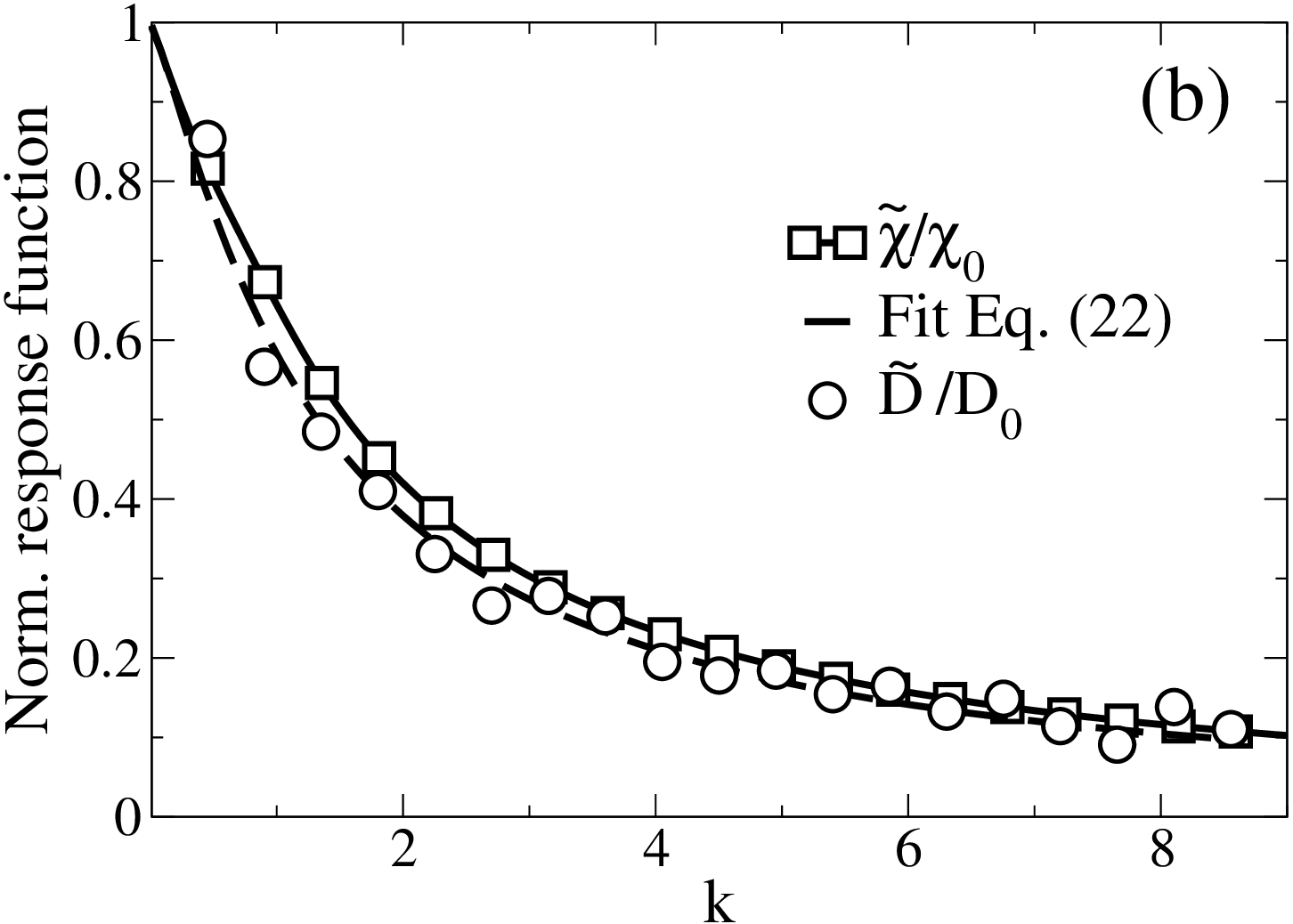}
  \caption{\label{fig:chiMolten} Molten salt (a) The Fourier
    components of $\chi$.  (b) Normalized kernels, $\w{\chi}/\chi_0$
    and $\w{D}/D_0$, for $T=T_\infty$.  }
\end{figure}

From Eq. (\ref{eq:endeq}) the Fourier components of $\chi$ can be
evaluated, the results is shown in Fig. \ref{fig:chiMolten} (a). The
evaluation is based on the fit of the diffusion kernel,
Eq. (\ref{eq:lorentz}), and the integral expression for the
charge-charge structure, Eq. (\ref{eq:SzzRadial}). First, for zero
screening, $T=T_\infty$, the response function is monotonically
decaying with respect to wavevector. This behavior is typically
observed for the diffusion and viscosity kernels
\cite{hansen:langmuir:2015}. For non-zero screening the response
features a maximum depending on temperature; the characteristic wave
length \color{red}$l=2\pi/k_{\text{max}}$ where $k_{\text{max}}$ is
the wavevector corresponding to maximum in $\widetilde{\chi}$,
\color{black} is approximately $l=2.5$ for $T=1.0177$ and $l=1.6$ for
$T=0.0177$. This means that application of an external field will
result in a relatively small flux, $j_i^e$, on large length scales and
a maximum for wavelength of roughly 2 atomic diameters. For
$T=0.0177$, we have that $\lim_{k\rightarrow 0} \w{\chi}(k)=0$ which
means that the charge density is zero at these length scales; this is
in agreement with perfect screening.
Another important point is that $\lim_{k\rightarrow \infty}S_{zz} =
1$, and if $\lim_{k\rightarrow \infty}D = 0$ as indicated in
Fig. \ref{fig:scatteringDkernel} we have that $\lim_{k\rightarrow \infty}
\w{\chi}=0$ according to Eq. (\ref{eq:endeq}).

In Fig. \ref{fig:chiMolten} (b) $\w{\chi}(k)/\chi_0$ and
$\w{D}(k)/D_0$ are depicted for the case $T=T_\infty$. The data show
good collapse, that is, there exists a master curve response
function. This identical wavevector dependence indicates that the
response functions are governed by the same underlying
process. Specifically, it is here conjectured that the $\chi$-response
is given by the diffusion processes in the system, i.e., cross
correlation effects can be ignored. \color{red} From
Fig. \ref{fig:chiMolten} (a) one can immediately see that this
collapse is not found \color{black} for the $T=1.0177$ and $T=0.0177$
cases, hence, different processes are involved.

\begin{figure}
  \includegraphics[scale=0.3]{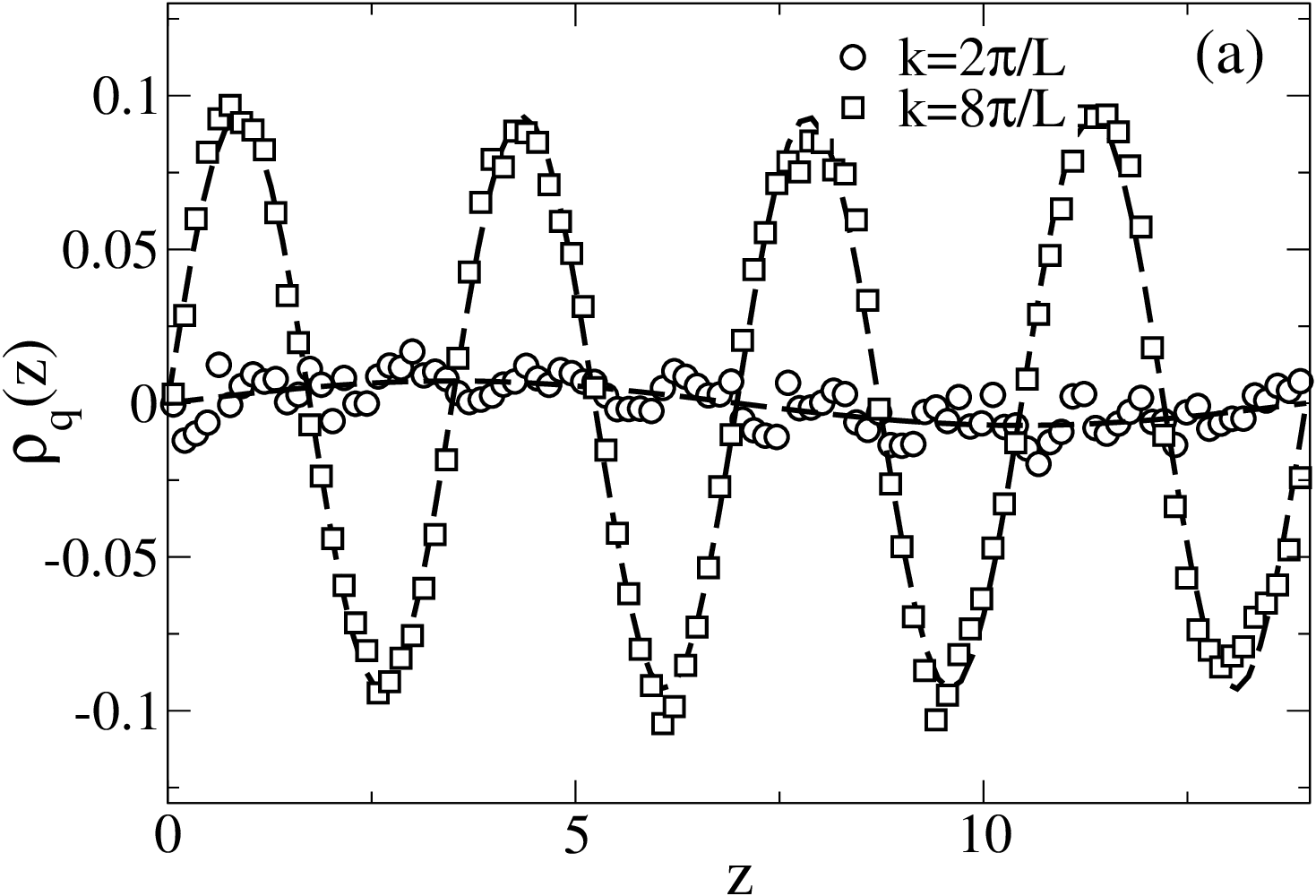}
  \includegraphics[scale=0.3]{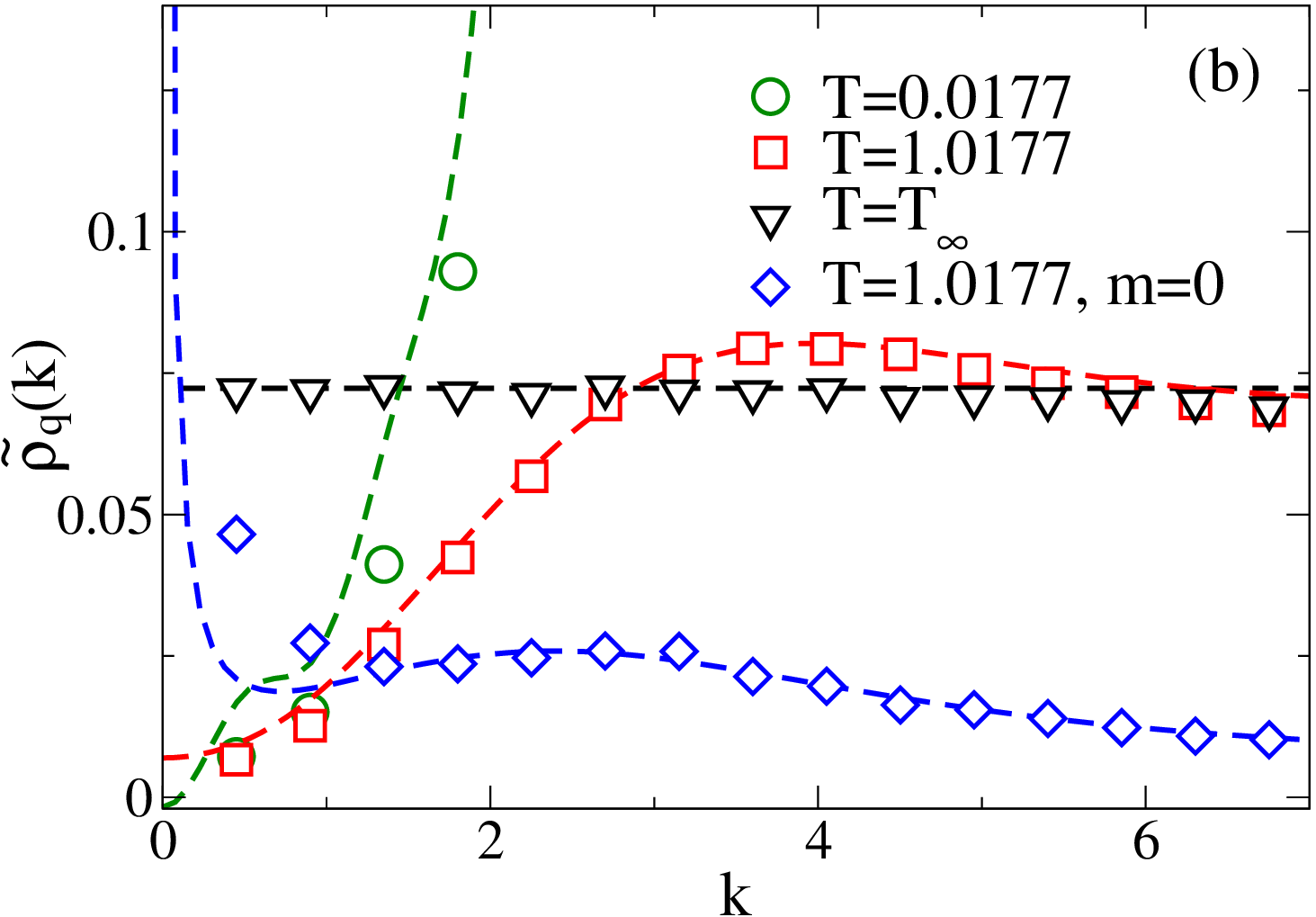}
  \caption{
    \label{fig:cmp} 
    [Color online] Non-equilibrium results for molten salt (a) Charge
    density profiles for $T=0.0177$. Lines are sine functions with
    amplitudes $\w{\rho}_q=0.0072$ and $0.093$, values obtained from a
    spectral analysis. (b) Charge density amplitudes for all three
    temperatures and for $m=0$. Symbols are simulation results. Lines
    are predictions from the theory, \color{red}
    Eq. (\ref{eq:FourierCompCharge}). \color{black} }
\end{figure}

The theory is compared with the non-equilibrium simulations. Figure
\ref{fig:cmp} (a) shows the charge density profile, $\rho_q$, for two
wavevectors $k=2\pi/L$ and $k=8\pi/L$ at $T=0.0177$. The system
length is $L=13.955$ and $m=1$, hence, the potential field amplitude
is constant. It is observed that the charge density amplitude is
larger for smaller wavelengths as expected. For $k>12\pi/L$ a simple
spectral analysis shows that higher order modes are excited
\color{red} compromissing Eq. (\ref{eq:FourierChargeSmall})
\color{black} and only results for $k<12\pi/L$ is shown.
Figure \ref{fig:cmp} (b) compares the amplitude for all three
temperatures with the predictions from the theory,
Eq. (\ref{eq:FourierCompCharge}). The agreement is excellent. Of
course, this comparison is equivalent to test the linear response,
Eq.(\ref{eq:linresponse}).  The case of $m=0$ is also shown, however,
the agreement is less satisfactory for low wavevectors, which is due to
the diverging amplitude in the limit of zero wavevector causing
a non-linear response and failure of the constitutive relation,
Eq. (\ref{eq:massflux}).

\subsection{Results: Ionic liquid}
\begin{figure}
  \includegraphics[scale=0.3]{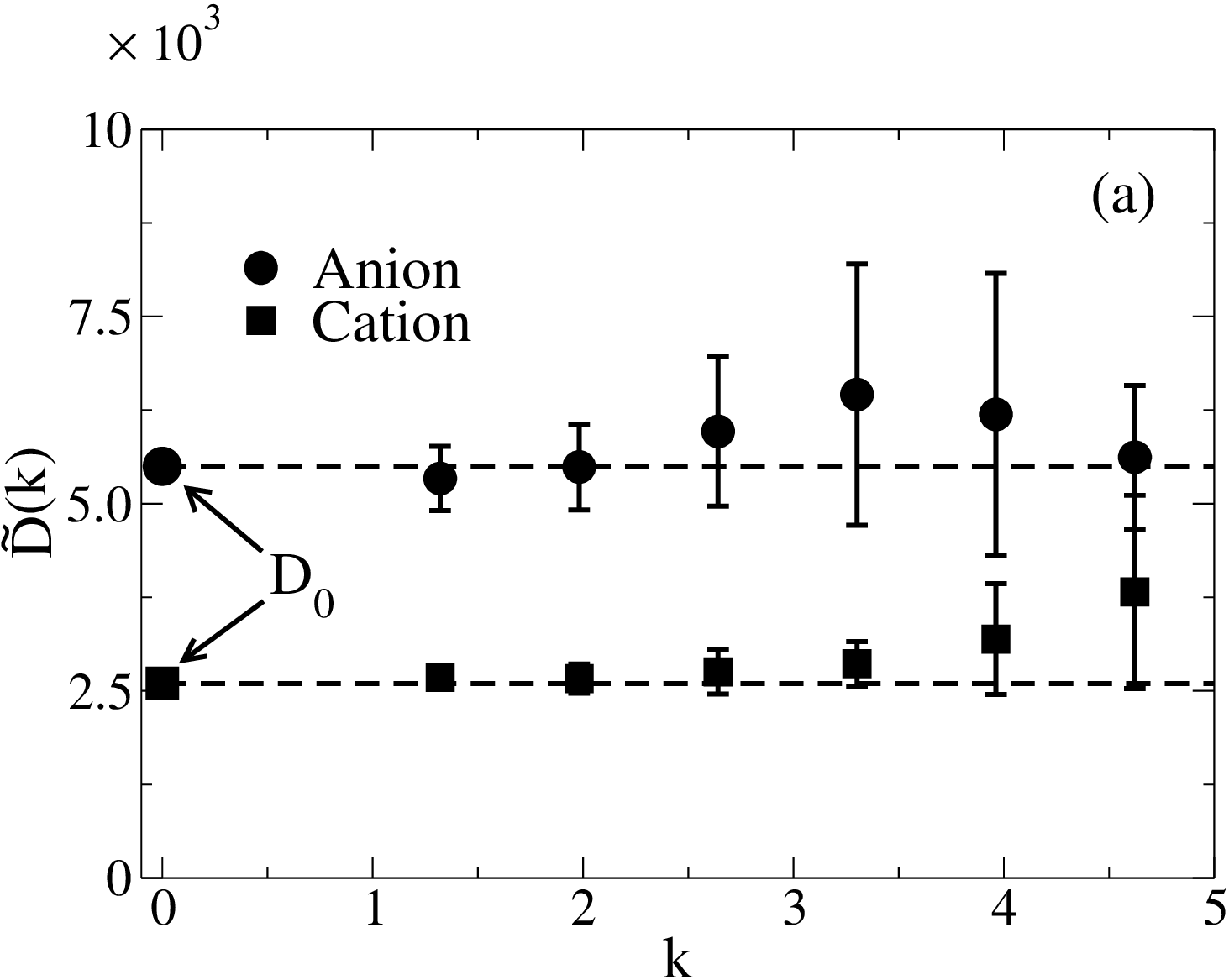}
  \\
  \vspace{1cm}
  \includegraphics[scale=0.3]{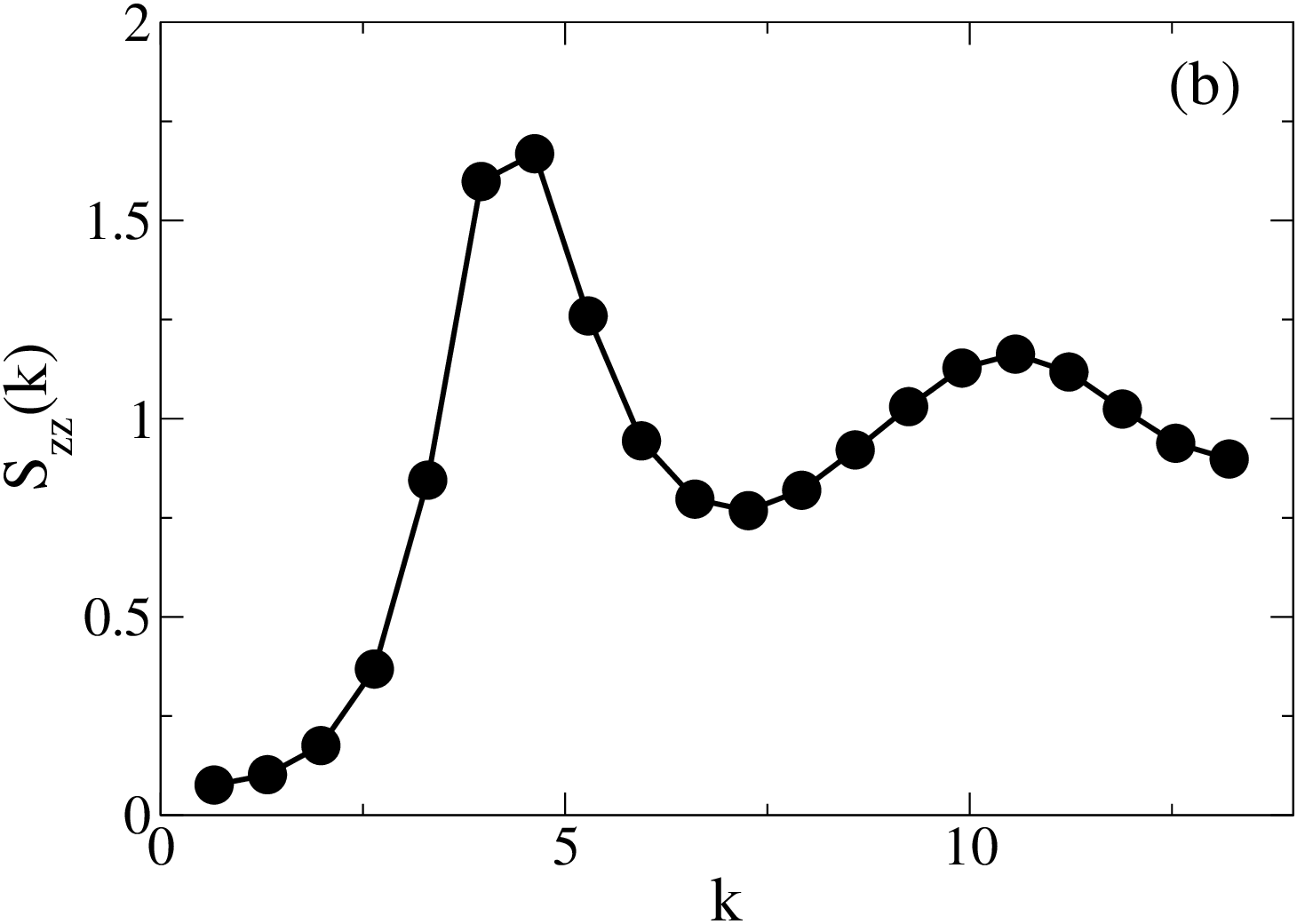}
  \\
  \vspace{1cm}
  \includegraphics[scale=0.3]{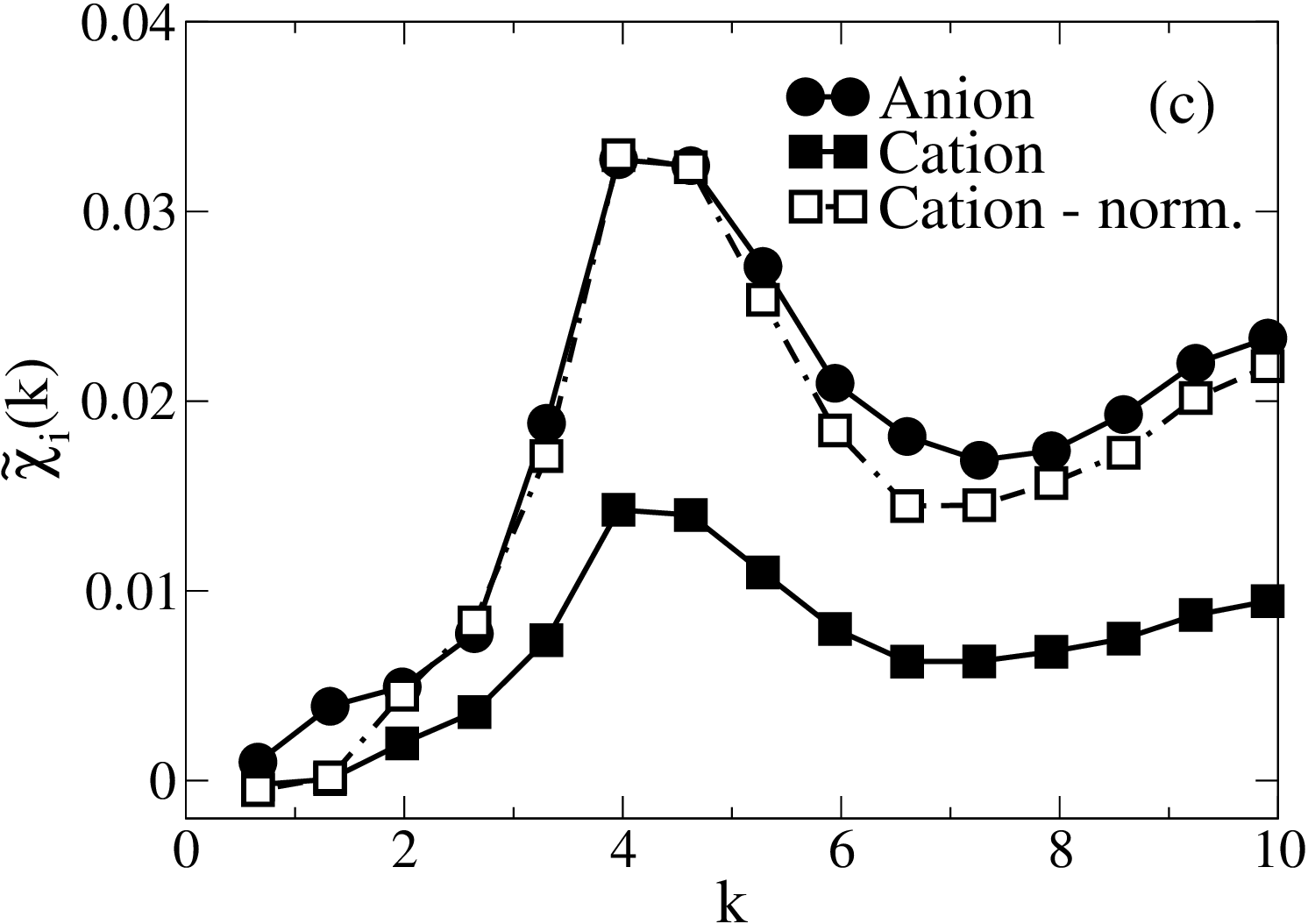}
  \caption{
    \label{fig:il1} 
    Ionic liquid. (a) Diffusion kernels for the anion and cation. (b)
    Charge-charge structure.  (c) wavevector dependent response
    function $\w{\chi}_i$ for the anion and cation. Squares connected
    with punctured line is the normalized cation response
    function. For all figures lines serve as a guide to the eye.  }
\end{figure}

In Fig. \ref{fig:il1} (a) the diffusion kernels are is shown for the
ion liquid model. These are evaluated as explained in
Sect. \ref{sec:moltensalt}. One sees that within statistical
uncertainty the diffusion kernel is wavevector independent, at least
up to $k=5$. Beyond this wavevector value the statistical error
increases dramatically and the results are non-conclusive. The
charge-charge structure, Fig. \ref{fig:il1} (b), is calculated from
the direct definition Eq. (\ref{eq:SzzDirect}). It features relatively
strong structure, that is, the system is in the screening regime. We
can therefore expect the response function $\chi$ to resemble low
temperature molten salt response function.

For the ionic liquid \color{red}
Eqs. (\ref{eq:FourierChargeSmall})-(\ref{eq:screening}) do not apply
as $\chi_+ \neq \chi_-$ and $D_+ \neq D_-$, and $\w{\chi}_i$ is found
from non-equilibrium simulations using Eq. (\ref{eq:fouriercomponets})
directly. This also means that we cannot compare the predictions from
these equations with simulation data. \color{black} The amplitudes of
the density profiles for both the anion and cation are analyzed giving
$\w{n}_i$. Note that only single modes are excited for the low
external field applied, $E_0=0.05$. Substitution of $\w{n}_i$ and
$\w{D}=D_0$ into Eq. (\ref{eq:fouriercomponets}) yields the results in
Fig. \ref{fig:il1} (c). The response features a maximum for $k\approx
4.25$ in good agreement with the maximum charge-charge structure.

To investigate if the two kernels can be mapped onto the same master
curve, the results from the cation kernel is normalized with respect
the maximum.  The normalized result is shown in Fig. \ref{fig:il1} (c) as
squares connected with a punctured line. To a reasonable agreement the
two kernels do follow a master curve which indicates that the
underlying mechanisms responsible for the response are the same. This
contrasts the wavevector independent diffusion kernel, that is, the
system response seen in the mass flux due to the density
gradient. Therefore, the physical mechanisms for the two fluxes
$j_i^d$ and $j_i^e$ are fundamentally different; at least in the
screening regime.

\section{Conclusion}
In this paper the \color{red} mass flux \color{black} of an ionic
system due to a spatially varying electric field is studied. Following
the Nernst-Planck equation two forces are present in this system: (i)
the concentration gradient and (ii) the gradient of the electric
potential. The two response functions (or kernels) that account for
the system response to these forces are the diffusion- and
$\chi$-response functions; \color{red}the $\chi$-response function
relates the mass flux with the external electric field excluding the
contribution from the concentration gradient (here modelled through
the self-diffusion). Note, this differs from the charge-charge
response function, $S_{zz}$, which relates the charge density to the
electric field including all underlying processes, and the ionic
conductivity that relates the charge current to the local
field.\color{black} In the limit of zero screening the $\chi$-response
function is directly related to the conductivity, on the other hand,
in the large screening regime the response function is related to the
charge-charge structure.

The spatial correlations in the system are manifested in the wavevector
dependence of the kernels. The molecular dynamics simulation data
show the diffusion and $\chi$-kernels feature very different
wavevector dependence in the screening regime. Interestingly, in the
screening regime both the molten salt and ionic liquid feature a
wavevector independent diffusion kernel and the response to the
external field is dominated by the charge-charge structure. This
latter quantity is a collective property of the system. In the
non-screening regime, on the other hand, the response to the external
field is closely related to the ionic conductivity and in this regime the
Nernst-Einstein relation holds to a good approximation, i.e.,
cross-correlation effects are negligible.

\bibliographystyle{unsrt}

\end{document}